# A Forecast for a South Heliopause Crossing by Voyager 2 in Late 2014 Using Intensity-time Features of Energetic Particles Observed by V1 and V2 in the North and South Heliosheaths

W.R. Webber[1] and D.S. Intriligator[2]

1.  New Mexico State University, Department of Astronomy, Las Cruces, NM  , 88003, USA

2. Carmel Research Center, Space Plasma Laboratory, Santa Monica, CA 90406, USA

## ABSTRACT

We have used corresponding "marker" features in the intensity vs. time profile of termination shock particles coming from within the heliosheath and also 5-12 MeV electrons which are coming from beyond the heliopause that are both observed at both V1 and V2 in the outer N and S heliosheaths, to estimate the thickness of the S heliosheath. These features in the N heliosheath have been observed on the CRS instrument on V1 in the N heliosheath beyond ~110 AU or within ~12 AU of the heliopause. Although the intensity-time features in the S heliosheath are quite different overall, certain features stand out at both V1 and V2. The correspondence of these features in the N and S heliosheath allows us to estimate that V2 will reach the S heliopause near the end of 2014.

**Introduction**

Voyager 1 has already traversed the heliosheath between 94.1 to 121.7 AU at an average N latitude ~32°. The intensity-time profiles of several different species of energetic particles observed by the CRS instrument (Stone, et al., 1977) have been reported earlier by Webber and Intriligator, 2013. These radial intensity profiles are relatively uniform over a large part of the N heliosheath; however, several features or structures in intensity vs. time profiles have been observed by V1, mainly for electrons, within about 10 AU of the heliopause (HP). These structures and their radial location in the heliosheath may be influenced by the local interstellar magnetic field, ISMF, and how it interacts with the heliospheric plasma. For example, the 1[st] of these structures in the outer N heliosheath, seen by V1 at a distance ~111 AU (Webber, et al., 2012), is about 10.5 AU inside the HP located at 121.7 AU (Stone, et al., 2013), and already has a magnetic field magnitude ~0.3 nT or about 2/3 of the ISMF (Burlaga and Ness, 2010; Burlaga, et al., 2013). This observed field is, energetically, 10-100 times stronger than a possible residual solar B field at that location. The large field amplitude therefore more likely has its magnitude influenced in some way by the ISMF at this location.

In comparison, as of the end of 2014, V2 will have traversed the heliosheath from 83.7 to 107 AU at an average S latitude ~29°. At this time V2 is therefore at a distance ~23.3 AU beyond the heliospheric termination shock (HTS). In the N hemisphere the heliopause was observed at a distance ~27.6 AU beyond the V1 HTS using the values for the HTS and HP given at the beginning of this section.

The intensity-time profiles of energetic particles are much more variable in the S heliosheath, particularly in the first 5 AU after the HTS crossing. These time variations have been recently discussed in Webber and Intriligator, 2013, and references therein. In late 2010, at a distance between 10-11 AU beyond the HTS a sustained large intensity increase begins at V2 for all energetic particle components. The peaks of these increases are observed at about 2011.3 at a distance of about 95 AU. At this time the TSP and ACR intensities are higher than those observed by V1 at any time during its heliosheath traversal.

After about 2011.3 all of the intensities at V2 then slowly decrease over the next 5.5 AU of outward radial movement of this spacecraft, so that by the end of 2012, when V2 is at 100.5 AU, the intensities of all components are at minima during which the intensities are not much larger than those observed when V2 crossed the HTS. A second intensity minimum, which is particularly noticeable for TSP, is observed about 0.9 year later near the end of 2013. These minima for TSP are not observed in the N hemisphere at any time by V1 and will be discussed in a separate paper.

If we first consider a nominal N heliosheath thickness of 27.6 AU corresponding to a heliopause at 121.7 AU and an HTS at 94.1 AU as observed by V1 and note that the relative distances of the N and S HTS crossings by V1 and V2 was equal to (94.1/83.7 AU) = 1.125. Then we assume the same ratio for the HP distances as for the HTS distances we obtain for the S HP distance, 27.6/1.125 ~24.3 AU. This assumes that any differences in the relative solar wind pressure at the times of the two HTS crossings are small and therefore that the factor of 1.125 is attributable mainly to external factors, dominated by the ISMF field and its direction. Thus at the end of 2014, when V2 is 23.3 AU beyond the HTS it is within about 1 AU of the South HP using this approach.

In this paper we will examine certain corresponding features in the N and S heliosheath structures in more detail which leads us to suggest that V2 may, in fact, reach the HP sometime in late 2014.

**Data**

The V1 and V2 data used here are shown in three figures. Figures 1 and 3 are similar in that the V1 and V2 times are synchronized at the time of the V1 and V2 crossings of the HTS which occurred 2.70 years later at V2. Thus we have delayed the V1 time by 2.70 years. This comparison is useful because V1 is moving outward at the rate of 3.62 AU/year, whereas for V2 this speed is 3.14 AU/year. The ratio of these outward speeds is therefore 1.145. This is almost the same as the ratio of the N-S asymmetry measured for the N-S HTS locations which is 1.125. This means that if the N and S heliosheath radial dimensions scale according to the time beyond the HTS crossing then the various features in each heliosheath, including the HP, will occur at the same relative "time" for V1 and V2 in each figure. Since the N heliosheath thickness is 27.6 AU the "nominal" S heliosheath thickness for this example would be 27.6/1.125=24.5 AU. If, however, corresponding features are observed in the N-S heliosheaths, and if they occur earlier in the S heliosheath, then the S heliosheath thickness might be compressed relative to the "nominal" N thickness, or visa-versa.

Figure 1 shows the data for 5-12 MeV GCR electrons. The radial profiles and relative intensities of this component in the outer N and S heliosheaths are greatly different. The higher latitude for V1, just above the HCS, may explain some of these N-S differences. The black circle on the V1 radial profile in Figure 1 occurs 10.5 AU inside the N heliopause. At this time at V1, a sudden jump in electron intensity occurs after a prolonged electron intensity increase at a rate ~10%/AU. At the same time the magnetic field amplitude increases up to a maximum >0.3 nT. As described by Webber, et al., 2012, this increase marks the entrance of V1 into a new region. In Figure 2 we show the V1 radial profiles of two different components (TSP and GCR electrons) in the N hemisphere from this time to the arrival of V1 at the heliopause about 3 years later. This history of V1 intensities was discussed by Webber, et al., in 2012, and Figure 2 here is an extension of Figure _ in that paper. Several new regions with different electron gradients and other features in the TSP and GCR nuclei were observed by V1 and their distances from the eventual heliopause crossing by V1 are indicated at the bottom of Figure 2.

Figure 3 shows the >0.5 MeV rate at both V1 and V2. This rate is dominated by ~1-10 MeV TSP. The detailed radial profiles in the outer N and S heliosheaths for these 1-10 MeV particles are again very different, however both have a distinct maximum which occurs at V1 in the N heliosheath about 3 AU inside the HP. The maximum in the V2 intensity-time profile for this component in the S heliosheath is shown by the red open circle ②.

In the N heliosheath, this increase is shown with a black circle 2 in Figure 3 for TSP. It occurs at a time when V1 was only ~3.0 AU in front of the N heliopause. This N heliosheath increase is believed to correspond to the increase labelled with a red circle ② in the intensity time-profile of TSP in the S heliosheath which occurred ~0.4 $\pm$ 0.2 year earlier on this scale. Both of these N and S increases are followed by a period of decreasing intensity for TSP and which is still continuing at V2.

## Discussion of Relative N-S Profiles

We wish to concentrate here on two features:  (1) The feature labelled ① in Figure 1 at both V1 and V2 in the intensity-time profiles of 5-12 MeV electrons.  This feature occurred when V1 was ~10.5 AU inside the heliopause; and in Figure 3, (2) the feature labelled ② at both V1 and V2 in the intensity time profiles of both TSP and GCR electrons and which occurred when V1 was within about 2.5 AU of the HP.

Consider first the feature ① in Figure 1.  Before this time both the N and S heliosheath electron intensities were increasing rapidly.  We believe this peak, occurring at 2009.5 in the N hemisphere and 2011.3 in the S hemisphere is a "corresponding" time marker in both hemispheres.  It occurred when V1 was 10.5 AU inside the N heliopause.  At V2 in the S heliosheath this peak occurs about 0.8 year earlier than at V1 in each of the intensity time profiles of GCR electrons and TSP.

Now consider Figure 3. At V1 another significant increase in TSP is seen at 2011.9, and is labelled ②.  This increase occurred when V1 was about 3.0 AU inside the eventual heliopause location.  It was followed by an intensity decrease of TSP up to the time where V1 was only ~0.3 AU inside the HP.  A similar intensity decrease in TSP is occurring at V2 in the S heliosheath.

This second red circle ② at the maximum intensity of TSP in the V2 profile in Figure 3 is therefore taken to be a "corresponding" increase of TSP in the S heliosheath.  It begins 0.5 ± 0.1 year earlier in the S heliosheath than in the N heliosheath.

## Summary and Conclusions

We have followed the intensity variations of the TSP >0.5 MeV as well as the GCR electrons between 5-12 MeV at V1 and V2 as these two spacecraft pass through the outer heliosheath.  One objective of this study is to determine when V2 might cross the heliopause using the V1 data as a guide.  We believe that V1 first crossed into a new region at about 2009.7, when it was 10.5 AU inside the HP (Webber, et al., 2010).  At that time the 5-12 MeV electron intensities and radial gradients changed dramatically along with the magnetic field energy density which increased by a factor >10, with the amplitude exceeding 0.3 nT, or over 2/3 of that of the ISMF measured beyond the heliopause.  This is the first reference time at V1.  We believe that this corresponding event is seen for electrons at V2 approximately 0.7 year earlier than at V1.

The next significant feature at V1 in the N heliosphere occurred about 2.4 years later, at about 2011.9, when V1 was at 118 AU, when a sharp peak was noted in the TSP intensity time profile. This peak occurred when V1 was between 2.5-3.0 inside the HP. This increase is labelled② in Figure 3. We use this event at V1 as a second "marker" for the progress of V2 through the heliosheath. At about 2014.1, the TSP intensity suddenly increased at V2, and then decreased up to the present time, 2014.6. This mimics the second "marker" event at V1 when it was 2.5-3.0 AU inside the HP. The increase at V2 occurred ~0.5 ± 0.1 year earlier than at V1 in our reference time scale.

Note that, in the reference time scale in Figures 1 and 3, V1 reaches the HP at 2015.35. If the S heliosheath were the same thickness as the N heliosheath, V2 would then reach the HP at the same time, when it was ~24.5 AU beyond the HTS. However, the two corresponding events described in this paper are observed earlier in the S heliosheath indicating an additionally compressed heliosheath. If we take the time for the final marker event ② which occurs ~0.5 year early at V2, then excluding possible further N and S heliosheath differences, the HP itself will occur ~0.5 ± 0.1 year early and will be crossed by V2 between 2014.75 and 2014.95.

Recent V2 data on 5-12 MeV GCR electrons shows an increase ~25%, starting in late May, 2014 and still continuing. We believe that this increase, labelled ③ in Figure 1, corresponds to the increase ~30% of electrons seen at V1 that starts in early May, 2012 and is labelled ③ in Figure 2. This increase started when V1 was 1.1 AU inside the HP which was crossed on August 25[th]. During this same time interval the TSP intensity continues to decrease in both the N and S heliosheaths. So this looks similar to the start of what was ultimately the final sequence of events at V1.

**Acknowledgements:** The authors thank JPL and the Voyager project director Ed Stone for their unwavering support during this 40 plus years mission. The ideas behind this study come from an earlier analysis of V1 and V2 CRS intensity time profiles of various particles in the N and S heliosheaths (Webber and Intriligator, 2013). In that paper we used plasma and magnetic field data available on the Web (John Richardson and Len Burlaga). We are also grateful for the support of the members of our CRS team N. Lal, A.C. Cummings and B. Heikkila, and the data available at the CRS public web-site (http://voyager.gsfc.nasa.gov). The work by D.S. Intriligator is supported by Carmel Research Center, Inc.

## REFERENCES

Burlaga, L.F. and N.F. Ness, (2010), Sectors and Large Scale Magnetic Field Strength Fluctuations in the Heliosheath Near 110 AU: Voyager 1-2009, Ap.J., 725, 1306-1316

Burlaga, L.F. and N.F. Ness, (2012), AGU Fall Meeting, Abstract SH-148-01,

Burlaga, L.F., N.F. Ness and E.C. Stone, et al., (2013), Magnetic Field Observations as Voyager 1 Entered the Heliosheath Depletion Region, Science, 341, 147-150

Stone, E.C., et al., (1977), Cosmic ray investigation for the Voyager missions: Energetic particle studies in the outer heliosphere – and beyond, Space Sci. Rev., 21, 335-376

Stone, E.C., et al., (2013), Voyager 1 observes low-energy galactic cosmic rays in a region depleted of heliospheric ions, Science, 341, 150-153

Webber, W.R. and D.R. Intriligator, (2013), Voyager 1 and 2 Observations in the North-South Heliosheaths and Implications for the Latitude extent of the Heliospheric Current Sheet and Sector Structure in the heliosheath, http://arXiv.org/abs/1403.332

Webber, W.R., et al., (2012), Sudden intensity increases and radial gradient changes of cosmic ray MeV electrons and protons observed at Voyager 1 beyond 111 AU in the heliosheath, GRL, 39, L06107

# FIGURE CAPTIONS

**Figure 1:** Intensities of low rigidity GCR electrons (5-12 MeV). Note the different radial profiles at V1 and V2 in the N and S heliosheaths, respectively. The RH scale for V1 GCR electrons is different than the LH scale for V2. Note also the sudden jump of electrons at 2009.7 at V1 when it was 10.5 AU inside the HP. We believe V1 entered a new region then (Webber, et al., 2012) and this point of reference is shown as a black circle ①. A corresponding S heliosheath point of reference is shown by the red circle ①, which is at the intensity peak at V2 after a sustained increase of a factor of 5. This increase factor and the two intensities at maximum (left and right hand axis) are roughly the same at this time.

**Figure 2:** V1 intensity-time profile of two components in the N heliosheath after the 2009.7 electron event at V1. The time at V1 is shown on the X-Axis scale which also shows the scale in AU inside the N heliopause. The events ①, ② and ③ are discussed in the text.

**Figure 3:** V1 and V2 intensities of >0.5 MeV particles (mainly ~1-10 MeV TSP). The events ① and ② in the N and S heliosheaths are discussed in the text.

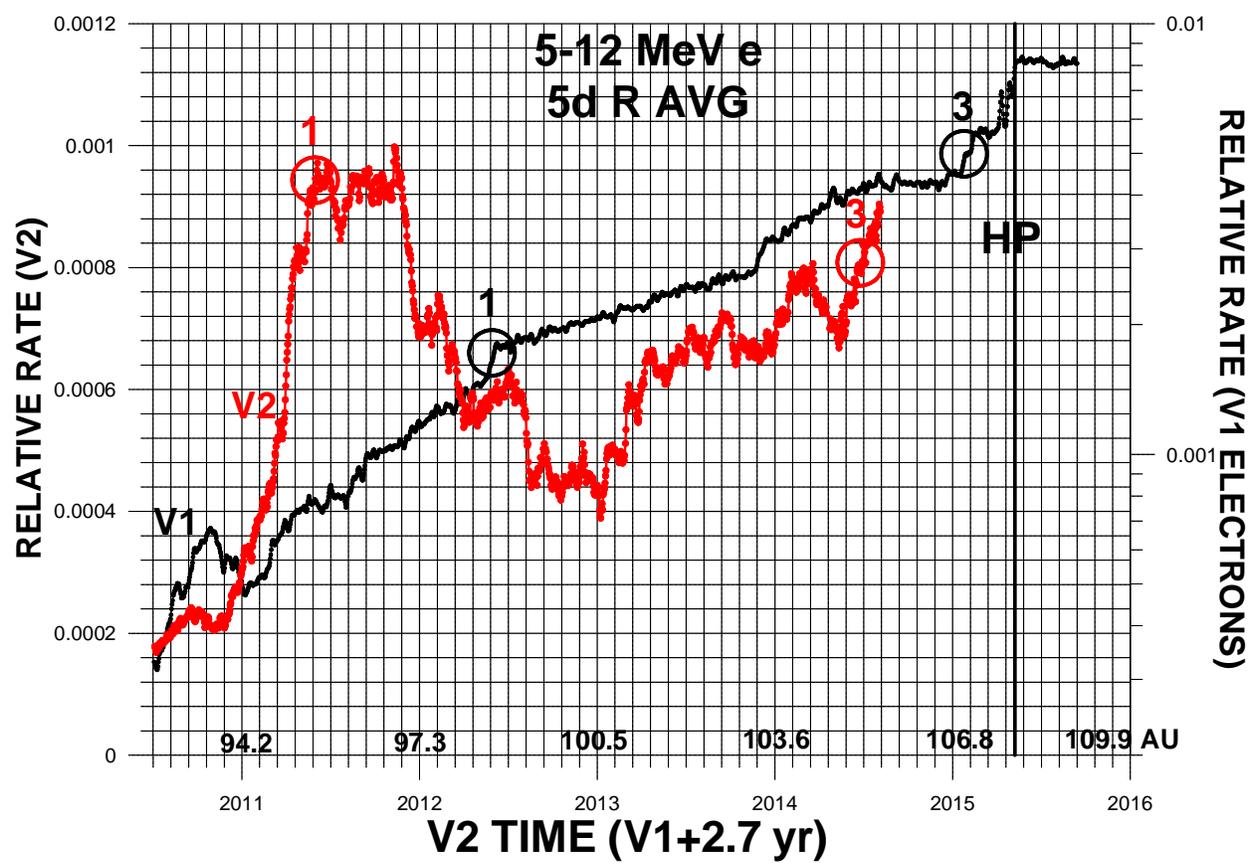

FIGURE 1

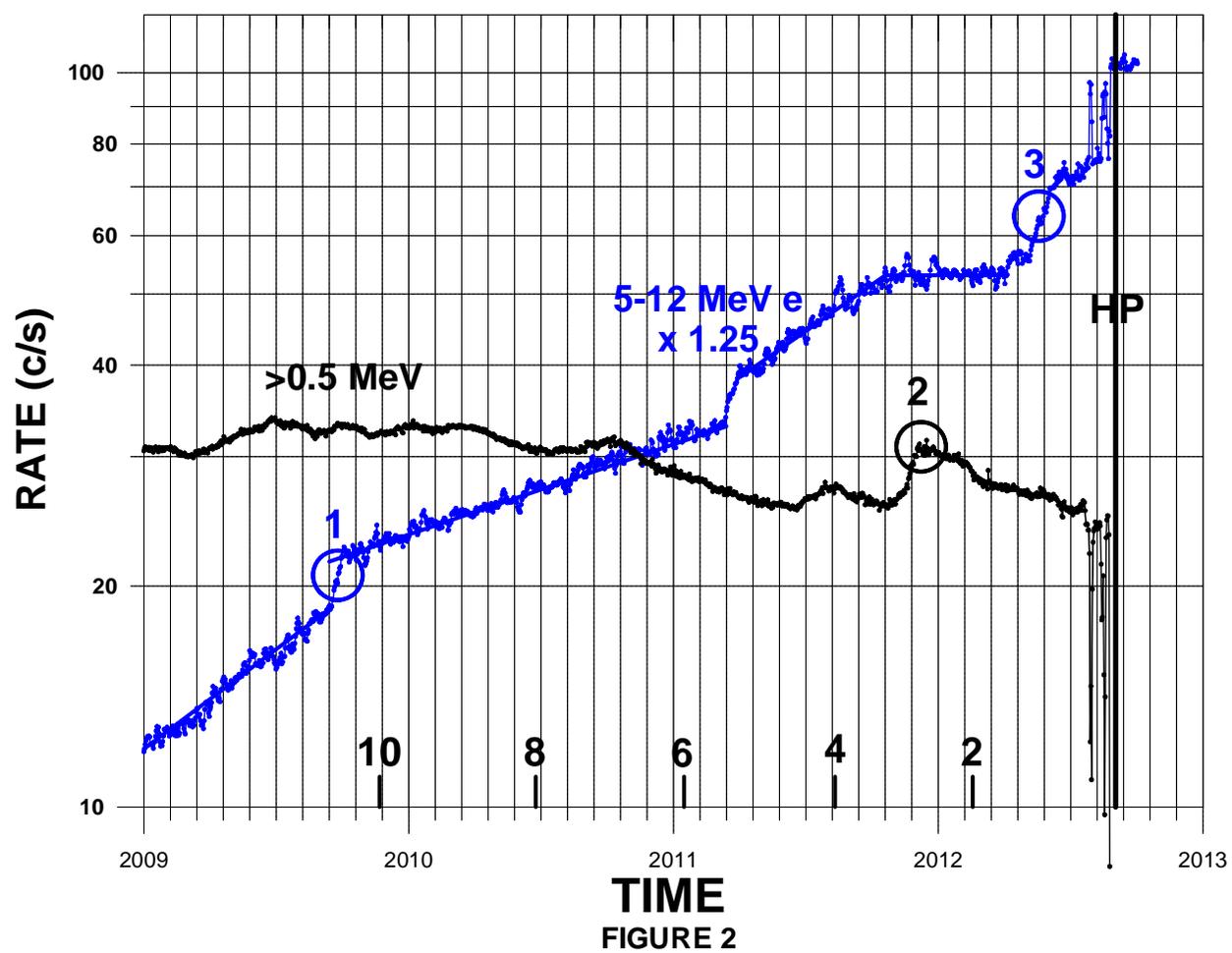

FIGURE 2

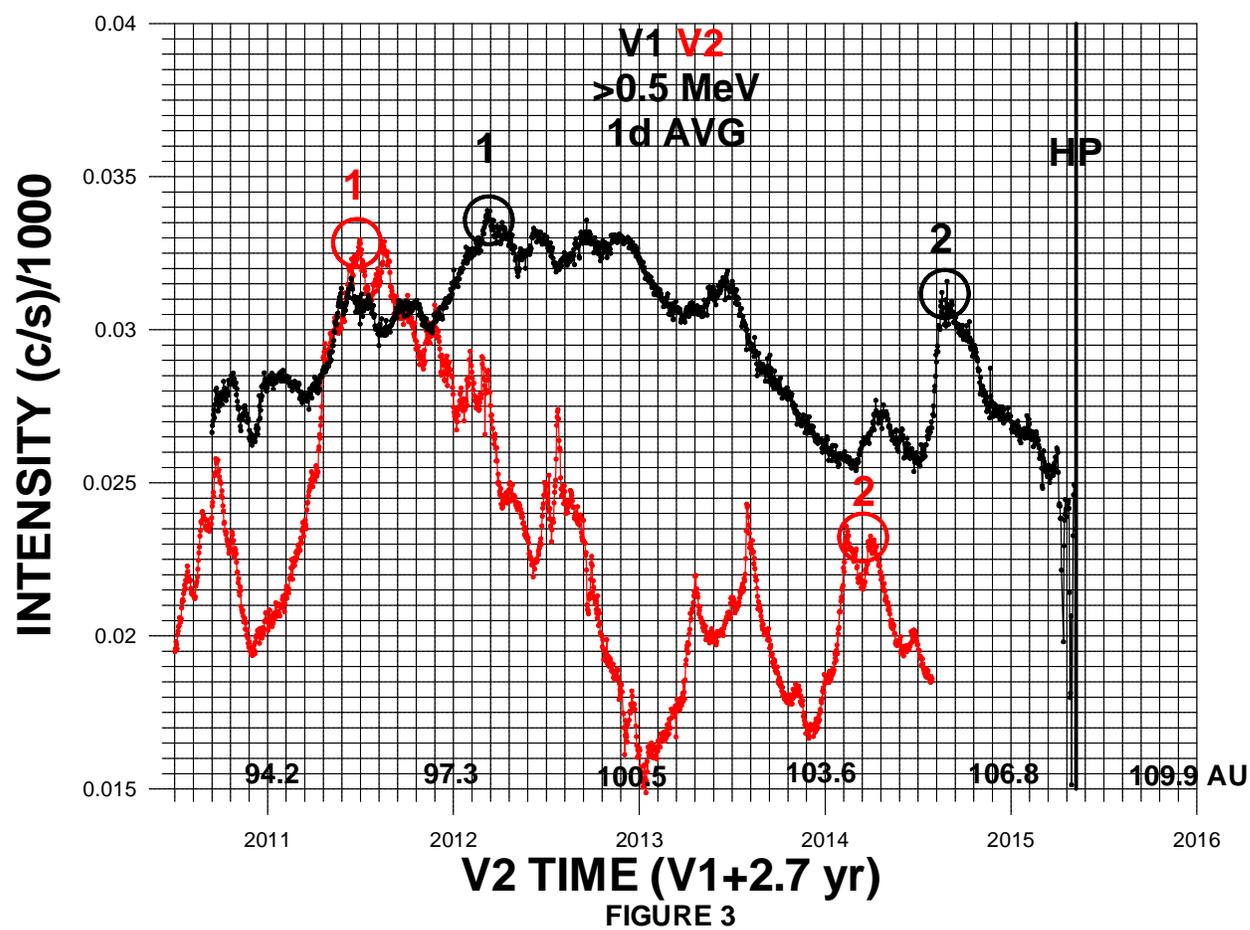

FIGURE 3